\documentstyle[12pt,aps]{revtex}

\widetext
\draft
\tighten
\begin{document}

\preprint{EFUAZ FT-96-33}

\title{Another Majorana Idea: \linebreak Real and Imaginary in
the Weinberg Theory\thanks{Submitted to ``Int. J. Theor. Phys."}}

\author{{\bf Valeri V. Dvoeglazov}}

\address{
Escuela de F\'{\i}sica, Universidad Aut\'onoma de Zacatecas \\
Antonio Doval\'{\i} Jaime\, s/n, Zacatecas 98068, ZAC., M\'exico\\
Internet address:  VALERI@CANTERA.REDUAZ.MX}

\date{September 8, 1996}

\maketitle

\bigskip

\begin{abstract}
The Majorana discernment of neutrality is applied to the
solutions of $j=1$ Weinberg equations in the $(j,0)\oplus (0,j)$
representation of the Poincar\`e group.
\end{abstract}


\newpage

It is well known that the Dirac equation can be separated in a relativistic
invariant way into a real part and an imaginary part~\cite{Maj}.
In the hamiltonian form the real Dirac equation reads:
\begin{equation}
\left [{1\over c}{\partial\over \partial t} - (\alpha,
\mbox{grad}) +\beta^\prime
\mu\right ] {\cal U} = 0\quad,
\end{equation}
where the Dirac matrices are chosen to be:
\begin{equation}
\alpha_x =\rho_1 \sigma_x ;\quad \alpha_y =\rho_3 ;\quad
\alpha_z =\rho_1 \sigma_z ; \quad
\beta^\prime = -i\beta =i\rho_1 \sigma_y\quad,\label{mm}
\end{equation}
and the mass term is $\mu=m c/ \hbar$. The anticommutaion rules are
in the configurational space
\begin{equation}
{\cal U}_i (q) \,
{\cal U}_k (q^\prime) + {\cal U}_k (q^\prime) \,
{\cal U}_i
(q) ={1\over 2} \delta_{ik} \delta (q - q^\prime)\quad\quad\quad.\quad
(Eq. 12^{\,\cite{Maj}})\nonumber
\end{equation}
The hamiltonian operator is then
defined
\begin{equation}
H = \int {\cal U}^\dagger \left [ -c(\alpha,p)
-\beta m c^2 \right ] {\cal U} dq\quad\quad\quad.\quad
(Eq. 13^{\,\cite{Maj}})\nonumber
\end{equation}
According to Majorana ``\ldots in
the present state of our knowledge equations (12) and (13) constitute the
simplest theoretical representation of a system of neutral particles".
Recent indications at the mass term of neutrino~\cite{ITEF} and at
the neutrino oscillations~\cite{LANL} induce one to look for a formalism
which could account a possible mass term and in the massless limit could
lead to the Weyl scheme~\cite{MLC,BZ,NP} in order to reproduce
some predictions of the standard model.

On the other hand, recent analysis of experimental data~\cite{tens}
in the decays of $\pi^-$ and $K^+$ mesons indicates
the necessity of introducing tensor interactions in theoretical models.
This induces one to correct our understanding the nature of the minimal
coupling and to pay attention to another types of Lorentz-invariant
interaction structures between the spinor and higher spin fields.
In the sixties the formalism was
proposed~\cite{Wein} for arbitrary spin-$j$  particles, which is
on an equal footing with the Dirac formalism  in the  $(1/2,0)\oplus
(0,1/2)$ representation. It allows for other forms of Lorentz-invariant
couplings~[8a,\S 7] and~\cite{DV-HJ1}. Moreover, it is on an equal
footing with the Dirac construct in the $(1/2,0)\oplus (0,1/2)$
representation and has no the problem of indefinite metric. The interest
in this description has been  considerably increased as a consequence of
constructing an explicit example of the theory of the
Wigner-type~\cite{BWW} and with the necessity of the detailed
interpretation of the $E=0$ solution of the Maxwell's
equations~\cite{Maj1,Oppen,Ahl1,Evans}.

In ref.~[6a] the author  proved that one cannot built
self/anti-self charge conjugate ``spinors" in the $(1,0)\oplus (0,1)$
representation. The equation $S^c_{[1]} \psi = e^{i\alpha} \psi$ has no
solutions in the field of complex numbers. The $\Gamma^5 S^c_{[1]}$
self/anti-self conjugate ``spinors" have been introduced there~\cite{NP}.
So, we have to look at alternative ways for describing  $j=1$ neutral
quantum fields. The aim of this note is to apply the
abovementioned Majorana idea of neutrality to the $j=1$ states of the
$(1,0)\oplus (0,1)$ representation.

In the generalized canonical (standard) representation the
Barut-Muzinich-Williams matrices are expressed:
\begin{mathletters}
\begin{eqnarray}
\gamma_{00}^{^{CR}} &=& \pmatrix{\openone&0\cr
0&-\openone\cr}\quad,\quad
\gamma_{i0}^{^{CR}}=\gamma_{0i}^{^{CR}} =\pmatrix{0&-J_i\cr
J_i&0\cr}\quad,\\
&\qquad& \gamma_{ij}^{^{CR}} = \gamma_{ji}^{^{CR}} =
\pmatrix{\eta_{ij} +\left \{ J_i, J_j\right \}&0\cr 0& -\eta_{ij} -\left
\{J_i,J_j\right \}\cr}\quad.
\end{eqnarray} \end{mathletters}
Here $J_i$,
\,\,$i,j=1,2,3$ are the $j=1$ matrices and
$\eta_{\mu\nu}$ is the flat space-time metric.
We work in the isotropic basis in which the spin matrices read
\begin{eqnarray}
J_x = {1\over \sqrt{2}}\pmatrix{0&1&0\cr
1&0&1\cr
0&1&0\cr}\quad,\quad
J_y ={i\over \sqrt{2}} \pmatrix{0&-1&0\cr
1&0&-1\cr
0&1&0\cr}\quad,\quad
J_z = \pmatrix{1&0&0\cr 0&0&0\cr
0&0&-1\cr}\quad.
\end{eqnarray}
By using the Wigner
time-reversal operator ($\Theta_{[j]} {\bf J} \Theta^{-1}_{[j]} =
-{\bf J}^\ast$)
\begin{equation}
\Theta_{[j=1]} = \pmatrix{0&0&1\cr
0&-1&0\cr 1&0&0\cr}
\end{equation}
one
can apply the Majorana procedure to transfer over the representation
where all $\gamma_{\mu\nu}$ matrices are the {\it real} matrices. The
unitary matrix ($U^\dagger U = U U^\dagger =\openone$) for this procedure
is
\begin{mathletters} \begin{eqnarray} U &=& {1\over
2\sqrt{2}}\pmatrix{(1-i) +(1+i) \Theta& -(1-i) +(1+i) \Theta\cr (1+i)
+(1-i)\Theta& -(1+i) + (1-i)\Theta\cr}\quad,\\
U^\dagger &=& {1\over
2\sqrt{2}}\pmatrix{(1+i)+(1-i)\Theta& (1-i)+(1+i)\Theta\cr -(1+i)
+(1-i)\Theta& -(1-i) +(1+i) \Theta\cr}\quad.
\end{eqnarray}
\end{mathletters}
As a result we arrive, $\gamma_{\mu\nu}^{^{MR}}=U\gamma_{\mu\nu}^{^{CR}}
U^\dagger$:
\begin{mathletters} \begin{eqnarray}
\gamma_{00}^{^{MR}} &=& \pmatrix{0&\Theta\cr \Theta
&0\cr}\quad,\quad
\gamma_{01}^{^{MR}} = \gamma_{10}^{^{MR}} =\pmatrix{0&-J_1
\Theta\cr -J_1 \Theta&0\cr}\quad,\\
\gamma_{02}^{^{MR}} &=& \gamma_{20}^{^{MR}} =\pmatrix{iJ_2 \Theta& 0\cr
0 & -iJ_2 \Theta\cr}\quad,\quad
\gamma_{03}^{^{MR}} = \gamma_{30}^{^{MR}} =\pmatrix{0&-J_3
\Theta\cr -J_3 \Theta&0\cr}\quad, \\
\gamma_{ij}^{^{MR}} &=& \gamma_{ji}^{^{MR}} = {1\over 2}
\pmatrix{i (J_{ij}^\ast -J_{ij} ) \Theta & (J_{ij}^\ast + J_{ij})\Theta\cr
(J_{ij}^\ast +J_{ij} )\Theta & -i (J_{ij}^\ast -J_{ij})
\Theta\cr}\quad\mbox{and}\quad
\gamma_5^{^{MR}} = \pmatrix{0&i\openone\cr
-i\openone & 0\cr}\quad.
\end{eqnarray} \end{mathletters} Here we introduced
the notation $J_{ij} = \eta_{ij} + \left \{ J_i, J_j \right \}$. Since
$J_2$ is the only (of the $J_i$) matrix which is imaginary in the
isotropic basis we can conclude that a set of real
Barut-Muzinich-Williams matrices is constructed.  The $(1,0)\oplus (0,1)$
functions in this representation are defined
\begin{mathletters}
\begin{eqnarray}
u^{^{MR}} ({\bf p}) &=& {1\over 2} \pmatrix{\phi_{_L} +
\Theta \phi_{_R}\cr \phi_{_L} +\Theta \phi_{_R}\cr} +{i\over 2}
\pmatrix{-\phi_{_L} +\Theta \phi_{_R}\cr \phi_{_L}
-\Theta\phi_{_R}\cr} = {\cal U}^+ +i{\cal V}^+\quad,\label{usp}\\
v^{^{MR}} ({\bf p}) &=& {1\over 2}
\pmatrix{-\phi_{_L} + \Theta \phi_{_R}\cr -\phi_{_L} +\Theta \phi_{_R}\cr}
+{i\over 2} \pmatrix{\phi_{_L} +\Theta \phi_{_R}\cr -\phi_{_L}
-\Theta\phi_{_R}\cr}= {\cal U}^- +i{\cal V}^-\quad. \label{vsp}
\end{eqnarray} \end{mathletters}
One can see that
\begin{equation}
v^{^{MR}} (p^\mu) = \gamma_5^{^{MR}}
u^{^{MR}} (p^\mu) = i\gamma_5^{^{WR}}\gamma_0^{^{WR}} u^{^{MR}} (p^\mu)
=\pmatrix{0&i\openone \cr -i\openone &0\cr} u^{^{MR}} (p^\mu)\quad.
\label{connect}
\end{equation}
The index {\it CR} stands for `the canonical representation', {\it
WR}, for `the Weyl representation', {\it MR}, for `the Majorana
representation'. For the second-type spinors~\cite{NP} $\lambda^{S,A}$
and $\rho^{S,A}$ in both the $j=1/2$ and $j=1$ case the use of the Majorana
representation leads to the natural separation into the real and
imaginary parts when referring to positive- (negative-) solutions.

The real and imaginary  parts of the positive-energy $u-$ bispinors of the
helicity $\pm 1$ are the following (cf.  with the $j=1/2$ case, see {\it
Appendix}):
\begin{mathletters}
\begin{eqnarray}
{\cal U}_\uparrow^+
(p^\mu) &=& {\cal U}_\downarrow^+ (p^\mu) = {1\over 2\sqrt{2}}\pmatrix{p^-
-{p_2 (p_1 +p_2)\over E+m}\cr -\sqrt{2} (p_1 -{p_2 p_3 \over E+m})\cr p^+
+{p_2 (p_1 -p_2) \over E+m}\cr p^- +{p_2 (p_1 -p_2)\over E+m}\cr -\sqrt{2}
(p_1 +{p_2 p_3 \over E+m})\cr p^+ - {p_2 (p_1 +p_2) \over
E+m}\cr}\quad,\quad\label{hp1a}\\
{\cal V}_\uparrow^+ (p^\mu) &=& -{\cal V}_\downarrow^+
(p^\mu) = {1\over 2\sqrt{2}}\pmatrix{-p^- +{p_1 (p_1 +p_2)\over E+m}\cr
-\sqrt{2} (p_2 +{p_1 p_3 \over E+m})\cr p^+ -{p_1 (p_1 -p_2) \over E+m}\cr
p^- -{p_1 (p_1 -p_2)\over E+m}\cr -\sqrt{2} (p_2 -{p_1 p_3 \over E+m})\cr
-p^+ + {p_1 (p_1 +p_2) \over E+m}\cr}\quad.\quad\label{hp1b}
\end{eqnarray}
\end{mathletters}
Surprisingly, real (and imaginary) parts of bispinors of different
helicities appear to be equal each other (within a sign). Thus, they are
connected by the operation of the complex conjugation. As to the solution
with $h=0$ one has only imaginary part of the positive-energy bispinor:
\begin{equation} {\cal U}_\rightarrow^+ (p^\mu) \equiv 0\quad,\quad {\cal
V}_\rightarrow^+ (p^\mu) = {1\over 2}\pmatrix{(p^- +m) {p_1+p_2 \over
E+m}\cr -\sqrt{2} (m+{p_1^2 + p_2^2\over E+m})\cr (p^+ +m) {p_1 - p_2\over
E+m}\cr (p^- +m) {p_2 - p_1\over E+m}\cr \sqrt{2} (m +{p_1^2 +p_2^2\over
E+m})\cr -(p^+ +m) {p_1 +p_2\over E+m}\cr}\quad.\label{h0} \end{equation}
The corresponding procedure can also be carried out for the
negative-energy solutions; the ``bispinors" are connected with
(\ref{hp1a},\ref{hp1b},\ref{h0}) using the equation (\ref{connect}).
Unlike to  `transversal' bispinors ($h=\pm 1$) the bispinor $v_\rightarrow
(p^\mu)$ has only a real part.

Finally, one  cannot find a matrix which transfers over the equation with
pure imaginary matrices because the unit matrix commutes with all
matrices of the unitary transformation.

In conclusion, using the standard form of the field operator in
the $x$- representation one can separate out the real and imaginary parts
of the $(1,0)\oplus (0,1)$ coordinate-space ``bispinors", then the result
can be compared with the case of the $(1/2,0)\oplus (0,1/2)$
representation. Relevant commutation relations can be found. However,
if we would wish to obtain an entirely real coordinate-space equation
(without any care of the $x-$ space imaginary part of the field function)
such a procedure leads to certain constraints between components of the
4-vector momentum and/or constraints on the phase factors. The physical
interpretation of the latter statement is unobvious and should be searched
in a separate paper.

{\it Acknowledgements.} I appreciate encouragments and discussions with
Profs. D. V. Ahluwalia, M. W. Evans, A.~F. Pashkov and G. Ziino. Many
internet communications from collegues are acknowledged.  I am grateful
to Zacatecas University for a professorship.  This work has been partly
supported by el Mexican Sistema Nacional de Investigadores, el
Programa de Apoyo a la Carrera Docente, and  by the CONACyT under the
research project 0270P-E.

\bigskip
\bigskip

{\tt Appendix.}  Here we wish to present the explicit forms of ${\cal
U}_{\uparrow\downarrow}^\pm$ and ${\cal V}_{\uparrow\downarrow}^\pm$, the
real and imaginary parts of bispinors in the $j=1/2$ Majorana
representation.  Comparing with the $j=1$ case we can observe differences.
The matrix of transfer over the Majorana representation from
the Weyl representation is
\begin{equation}
U ={1\over 2}\pmatrix{\openone -i\Theta & \openone +i\Theta\cr
-\openone -i\Theta & \openone -i\Theta\cr}\quad,\quad
U^\dagger = {1\over 2}\pmatrix{\openone -i\Theta & -\openone -i\Theta\cr
\openone + i\Theta & \openone -i\Theta\cr}\quad.
\end{equation}
The $\gamma$- matrices are given by
\begin{mathletters}
\begin{eqnarray}
&&\gamma_{_{MR}}^0 = \pmatrix{0& -i\Theta_{[1/2]}\cr
-i\Theta_{[1/2]} & 0\cr}\quad,\quad
\gamma_{_{MR}}^1 = \pmatrix{-i\sigma_1 \Theta_{[1/2]} & 0\cr
0& -i\sigma_1 \Theta_{[1/2]}\cr} \\
&& \gamma_{_{MR}}^2 = \pmatrix{0& -\sigma_2\cr
\sigma_2 &0 \cr}\quad,\quad
\gamma_{_{MR}}^3 =\pmatrix{-i\sigma_3 \Theta_{[1/2]} & 0\cr
0& -i\sigma_3 \Theta_{[1/2]}\cr}\quad,\quad\\
&&\mbox{and}\quad
\gamma_{_{MR}}^5 = \pmatrix{-i\Theta_{[1/2]} &0\cr
0& i\Theta_{[1/2]}\cr}\quad .
\end{eqnarray}
\end{mathletters}
All they are imaginary and are related with Eq. (\ref{mm}).
\begin{mathletters}
\begin{eqnarray}
{\cal U}^+_\uparrow (p^\mu) &=&  {1\over
2\sqrt{(E+m)}}\pmatrix{E+m -p_2 \cr 0\cr -p_3 \cr -
p_1\cr}\quad,\quad
{\cal U}^+_\downarrow (p^\mu) = {1\over
2\sqrt{(E+m)}}\pmatrix{0\cr E+m -p_2 \cr -p_1\cr p_3
\cr}\,\, ,\\
{\cal V}^+_\uparrow (p^\mu) &=&
\gamma_5^{^{WR}} \gamma_0^{^{WR}} {\cal U}_\downarrow^+ (\widetilde
p^{\,\mu}) = {1\over 2\sqrt{(E+m)}}\pmatrix{p_1\cr -p_3 \cr 0\cr -E -m -
p_2 \cr}\quad, \quad\\
{\cal V}^+_\downarrow (p^\mu) &=& -\gamma_5^{^{WR}}
\gamma_0^{^{WR}} {\cal U}_\uparrow^+ (\widetilde p^{\,\mu}) ={1\over
2\sqrt{(E+m)}}\pmatrix{-p_3 \cr -p_1\cr E+m +p_2\cr 0\cr}\quad.
\end{eqnarray} \end{mathletters}
The negative-energy spinors are related
with the positive-energy ones by using the formulas:
\begin{equation}
v^{^{MR}}_\uparrow (p^\mu) = -i [ u^{^{MR}}_\downarrow
(p^\mu) ]^\ast\quad,\quad v^{^{MR}}_\downarrow (p^\mu) =
+ i [ u^{^{MR}}_\uparrow (p^\mu) ]^\ast \quad,
\end{equation} and, thus,
\begin{mathletters}
\begin{eqnarray}
{\cal U}^+_\uparrow &=& {\cal
V}^-_\downarrow = \Re e \,u_\uparrow^{^{MR}} = {u_\uparrow^{^{MR}} -
iv_\downarrow^{^{MR}} \over 2}\quad,\quad {\cal U}^+_\downarrow = -{\cal
V}^-_\uparrow = \Re e \, u_\downarrow^{^{MR}} = {u_\downarrow^{^{MR}} +
iv_\uparrow^{^{MR}} \over 2}\quad,\quad \\ {\cal V}^+_\uparrow &=& {\cal
U}^-_\downarrow = \Im m\, u_\uparrow^{^{MR}} = {u_\uparrow^{^{MR}}
+iv_\downarrow^{^{MR}} \over 2i}\quad,\quad {\cal V}^+_\downarrow = -
{\cal U}^-_\uparrow = \Im m\,u_\downarrow^{^{MR}} = {u_\downarrow^{^{MR}}
-iv_\uparrow^{^{MR}} \over 2i}\quad.
\end{eqnarray} \end{mathletters}

These formulas also can be used to form  even- and odd- bispinors
with respect to ${\bf p} \rightarrow -{\bf p}$.

\end{document}